\documentclass[aoas,nameyear,seceqn,dvips]{arximspdf}
\usepackage{dcolumn,multirow}
\usepackage{graphicx}

\doi{10.1214/09-AOAS319}
\volume{4}
\issue{3}
\pubyear{2010}
\firstpage{1602}
\lastpage{1620}

\makeatletter
\newtheorem{theorem}{Theorem}[section]
\newcolumntype{d}[1]{D{.}{.}{#1}}
\makeatother

\begin{document}
\begin{frontmatter}

\title{Backward estimation of stochastic processes with failure events as time origins\protect\thanksref{T1}}
\pdftitle{Backward estimation of stochastic processes with failure events as time origins}
\runtitle{Backward estimation of stochastic processes}
\thankstext{T1}{Supported in part by the National Institutes of Health Grant
P01 CA 098252.}

\begin{aug}
\author[A]{\fnms{Kwun Chuen Gary} \snm{Chan}\ead[label=e1]{kcgchan@u.washington.edu}\corref{}}
and
\author[B]{\fnms{Mei-Cheng} \snm{Wang}\ead[label=e2]{mcwang@jhsph.edu}}

\runauthor{K. C. G. Chan and M.-C. Wang}

\affiliation{University of Washington and Johns Hopkins University}

\address[A]{Department of Biostatistics\\
University of Washington\\
Seattle, Washington 98185\\
USA\\
\printead{e1}}

\address[B]{Department of Biostatistics\\
Johns Hopkins University\\
Baltimore, Maryland 21205\\
USA\\
\printead{e2}}
\end{aug}

\received{\smonth{9} \syear{2009}}

\begin{abstract}
Stochastic processes often exhibit sudden systematic changes in pattern
a short time before certain failure events.  Examples include increase
in medical costs before death and decrease in CD4 counts before AIDS
diagnosis.  To study such terminal behavior of stochastic processes, a
natural and direct way is to align the processes using failure events
as time origins.  This paper studies backward stochastic processes
counting time backward from failure events, and proposes one-sample
nonparametric estimation of the mean of backward processes when
follow-up is subject to left truncation and right censoring.  We will
discuss benefits of including prevalent cohort data to enlarge the
identifiable region and large sample properties of the proposed
estimator with related extensions. A SEER--Medicare linked data set is
used to illustrate the proposed methodologies.
\end{abstract}


\end{frontmatter}

\section{Introduction}\label{sec1}

Stochastic processes such as recurrent events and repeated measurements
are often collected in medical follow-up studies in addition to
survival data. Examples include recurrent hospitalizations,  medical
cost processes, repeated quality of life measurements and CD4 counts.
Such processes often exhibit certain terminal behavior during a short
time before failure events.  For example, medical costs tend to
increase suddenly before death, qualities of lives deteriorate before
death and CD4 counts decrease before AIDS diagnosis.

Conventional statistical methodologies mainly focus on stochastic
processes that are counting forward from initial events observed for
every individual; see Nelson (\citeyear{N1988}), Pepe and Cai (\citeyear{PC1993}), Lawless and
Nadeau (\citeyear{LN1995}), Cook and Lawless (\citeyear{CL1997}),  Lin et al. (\citeyear{Letal2000}) and Wang, Qin
and Chiang (\citeyear{WQC2001}), among others, on recurrent event processes, Lin
(\citeyear{L2000}) on medical cost processes and Pawitan and Self (\citeyear{PS1993}) on CD4
count processes.  The conventional views of stochastic processes,
however, are not designed to study the terminal behavior of processes.
Consider medical cost as an example.  Calculating the mean of cost
processes for a population defined at an initial event would include
both survivors and nonsurvivors at any fixed time after the initial
event, and the increase in medical cost based on survivors' cost
measurement is offset by nonsurvivors who do not contribute to the
increase in medical cost after death.  Unless the failure times are
constant over a population,  conventional forward processes do not
serve the purpose of estimating the terminal behavior of stochastic
processes.

In this paper we directly consider stochastic processes before failure
events of interest, by introducing backward processes that start at
failure events and counting backward in time.  By aligning the origins
of the processes to failure events, terminal behavior of stochastic
processes could be naturally and directly studied by the backward
processes.  We will focus on one-sample nonparametric estimation of the
mean of backward processes when the failure events are partially
observed subject to left truncation and right censoring.  Since failure
events and processes right before failure events may not be observed,
statistical methods are needed to correct a bias induced by
missingness.  Development of methods rely on a stochastic
representation technique of a marked counting process generalizing that
of Huang and Louis (\citeyear{HL1998}) and the proposed estimator also generalizes a
weighted estimator for left truncated and right censored data proposed
by Gross and Lai (\citeyear{GL1996}).

Throughout this paper we will consider medical costs as motivating examples.
The SEER--Medicare linked data provide illustrative examples of medical cost process data collected in a left truncated and right censored
follow-up sample.  The Surveillance, Epidemiology and End-Results
(SEER)--Medicare linked data are population-based data for studying
cancer epidemiology and quality of cancer-related health services.
The SEER--Medicare linked data consist of a linkage of two large
population-based databases, SEER and Medicare. The SEER data
contain information of cancer incidence diagnosed between 1973 and
2002.  The Medicare data contain information on medical costs
between 1986 and 2004. The linked data consist of
cancer patients in the SEER data who were enrolled in Medicare
during the study period of the Medicare data.  Details of each
data and linkage are discussed in Warren et al. (\citeyear{Wetal2002}).  Although
the linkage criterion sounds simple, it creates a left truncated and
right censored sample because the two data sets have different
starting times. In the SEER--Medicare linked data, patients diagnosed
with cancer before 1986 form a prevalent cohort, because only those
patients who survived through 1986 were included.  Patients
diagnosed with cancer after 1986 form an incident cohort, because
those patients were recruited at the onset of disease. Patients
survived through 2004 were considered censored. A prevalent cohort is
typically a left truncated and right censored sample and data from a
combination of incident and prevalent cohorts are also left
truncated and right censored.

This article is organized as follows.  In Section \ref{sec2} we will introduce
backward processes to study terminal behavior of stochastic processes
and discuss the differences  from conventional forward process models.
The proposed methods for estimating the mean function of backward
processes will be discussed in Section~\ref{sec3}, together with identifiability
problems associated with incomplete follow-up, large sample properties
of the proposed estimators and a method for constructing confidence
bands for mean functions. We will also discuss two related extensions
of the proposed procedure in Section \ref{sec4}, one is on distributional
estimation of the backward processes, and the other is on estimation of
derivatives of backward mean functions. Simulations and real examples
analyzing a SEER--Medicare linked data set will be presented in Section
\ref{sec5}. Section \ref{sec6} will include several concluding remarks.

\section{Forward and backward processes}\label{sec2}
\begin{figure}[b]

\includegraphics{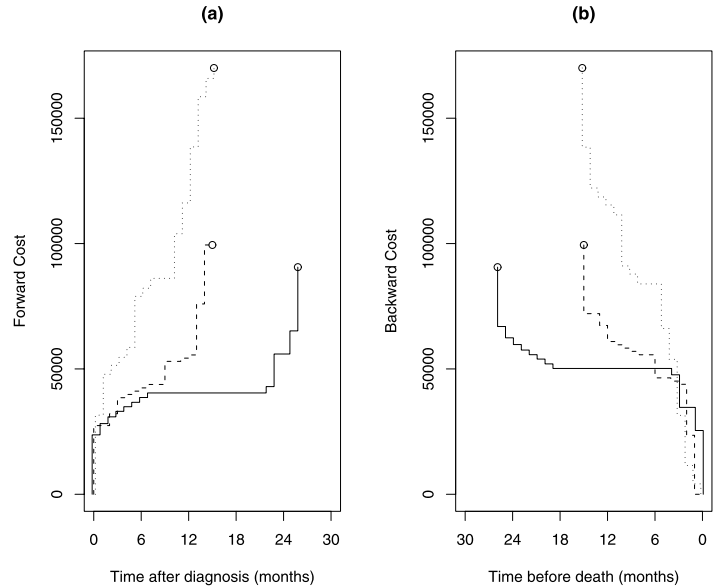}

\caption{Trajectories of forward and backward cost processes for 3
uncensored individuals in the SEER--Medicare linked data. \textup{(a)} Forward cost
processes.  Circles represent failure events. \textup{(b)} Backward cost
processes.  Circles represent diagnoses of cancer.}\label{fig1}
\end{figure}

Let $Y(t)$ be a stochastic process with bounded variation, where $t$ is
the time after an initial event, usually defined as the time of disease
onset. We call $Y(t)=\int_0^t dY(s)$ a forward stochastic process since
the time index $t$ in $Y(t)$ starts at the initial event and moves
forward with calendar time. On the other hand, a backward stochastic
process is defined as $V(u)=\int_{T-u}^TdY(s)$, where $T$ is the time
from the initial event to a failure event of interest and the time
origin for $V(u)$ is the failure event.  In the medical cost example,
$Y(t)$ is total medical cost within $t$ time units after the initial
event,  and $V(u)$ is total medical cost during the last $u$ time units
of life.  Figure \ref{fig1} shows the trajectories of forward and backward cost
processes for 3 uncensored individuals in the SEER--Medicare linked
data.

In Figure~\ref{fig1} we can see an increase in medical cost a short
period before death.  To study this terminal behavior of medical cost processes, it is natural to align the
processes to a different time origin, the failure event, as shown in
Figure~\ref{fig1}(b). Since terminal behavior of stochastic processes usually incur during a short time period before death, relevant scientific questions
center on a rather short period $\tau_0$ before death,
say, 6 months or 1 year. $\tau_0$ is a prespecified time period related to scientific questions of interest. The backward stochastic processes at $\tau_0$ time units before
failure events are only meaningfully defined for a subgroup of
patients who survive at least $\tau_0$ time units, and the estimand of
interest is $E(V(u)|T\geq \tau_0)$, for $u \in [0,\tau_0]$. However,
due to limited study duration, only a conditional version
$\mu_{\tau_0,\tau_1}(u)=E(V(u)|\tau_0\leq T<\tau_1)$ can be
nonparametrically identified, where $\tau_1$ depends on study design and data availability.  $\tau_1$ can be taken as the maximum follow-up period, and the time period of interest $\tau_0$ is usually much shorter than $\tau_1$.
We will further discuss implications of incident and prevalent
sampling on the identifiability constraints in Section \ref{sec3.2}.

To distinguish between processes with time origins at initial events
and failure events, throughout this paper $t$ denotes a time index
counting forward from initial events and $u$ denotes a time index
counting backward from failure events. The processes $Y(t)$ and $V(u)$
address different scientific questions and have different
interpretations. Consider the medical cost example, where $Y(t)$
measures medical cost from an initial event.  Note that $Y(t)$ will not
increase after death, so that $Y(t)=Y(T)$ for $t\geq T$.  The
interpretation of forward mean function $E(Y(t))$ is generally
confounded with survival performance. For example, if there are two
groups of patients with the same spending per unit time when alive but
different survival distributions, the group with longer survival time
will have a higher mean forward cumulative cost.  There may also be
crossovers between mean forward cost curves, because patients with
severe disease tend to spend more near disease onset but die in shorter
time than patients with less severe disease. We shall see such an
example from the SEER--Medicare data set in Section \ref{sec5.2}.  On the other
hand, the time origin of a backward process $V(u)$ is defined to be a
failure event, and the backward mean function can be interpreted as the
mean of stochastic processes before failure events. In the medical cost
example, when financial decision is a major concern (e.g., decision
made by insurance company), then discounted forward cost may be more
relevant.  The backward processes essentially answer different types of
questions related to end-of-life cost, and there is currently a lot of
public health interest in comparing and evaluating palliative care.
This work could provide valid statistical methods for public health
researchers interested in estimating end-of-life medical cost, together
with other applications.

\section{Proposed estimation}\label{sec3}
\subsection{Data structure}\label{sec3.1}
Let $T$ be a failure time, $C$ be a censoring time and $W$ be a
truncation time. $(T, C, W)$ are defined relative to an initial event.
Truncation time $W$ is the time between the initial event and the time
of recruitment. For incident cases, $W=0$.  For prevalent cases, $W>0$
and survival data are observed only when $T\geq W$, that is, the
failure time is left truncated. Also, since censoring is only
meaningfully defined for subjects who are eligible to be sampled, we
assume that $P(W\leq C)=1$ as discussed in Wang (\citeyear{W1991}).
Let $X=\min(T,C)$ and $\Delta=I(T\leq C)$. In addition to observing the
usual left truncated and right censored survival data $(W,X,\Delta)$,
$Y(t)$ is also observed from time of recruitment to death or censoring.
We assume an independent censoring and truncation condition in which
$\{V(\cdot),T\}$ is independent of $\{W,C\}$.  This assumption does not
impose any dependent structure between the process $V(\cdot)$ and the
failure time $T$.  In fact, $V(\cdot)$ and $T$ are allowed to be
arbitrarily dependent under this assumption and thus handle the case of
informative failure events.  The assumption is similar in nature to
those imposed for nonparametric estimation of forward mean function
with informative terminal events; see, for example, Lawless and Nadeau
(\citeyear{LN1995}), Lin et al. (\citeyear{Letal1997}), Strawderman (\citeyear{S2000}) and Ghosh and Lin (\citeyear{GL2000}).
Let $S(t)=P(T\geq t)$ and $G(t)=P(X\geq t \geq W|T\geq W)$, by the
independent censoring and truncation conditions $G(t)=S(t)\cdot P(C\geq
t\geq W)/\beta$ where $\beta=P(T\geq W)$.

To estimate the mean of $V(u)$ for $u\in[0,\tau_0]$, we only
need the following minimal data [Huang and Louis (\citeyear{HL1998})]:
\[
\bigl\{W_i,X_i,\Delta_i,\{\Delta_iV_i(u),u\in[0,\tau_0]\},i=1,\ldots,n\bigr\}.
\]
That is, in addition to the survival data, we only need backward
process data to be available for individuals whose failure events are
uncensored. For subjects in a prevalent cohort, backward process data
may not be fully available for individuals who experience failure
events within $\tau_0$ from recruitment.  In this case, we may treat
recruitment time to be $\tau_0$ after the actual recruitment date and
$W+\tau_0$ be a new truncation variable for the subjects in a prevalent
cohort.  This is equivalent to artificially truncating a small portion
of data and it guarantees that $V(u), u\in [0,\tau_0]$, is observable
for all uncensored observations with $T\geq \tau_0$ in the prevalent
cohort.

\subsection{Identifiability}\label{sec3.2}
A backward stochastic process can be viewed as a marked process attached
to a failure event.  This is a generalization of marked variables
considered by Huang and Louis (\citeyear{HL1998}) in which random variables are
observed at failure events.  Because of limited study duration,
marginal distribution of marked variables cannot be fully identified nonparametrically.
The same applies to backward stochastic processes because we do
not have data on backward processes for subjects with survival
time greater than $\tau_1$, which is the maximum support of the
censoring time.  In view of this identifiability problem, together
with the fact that stochastic processes within a prespecified time period of interest $\tau_0$ before failure events are only meaningfully defined for the subgroup of individuals who
survives at least $\tau_0$ time units, we confine ourselves to
estimate a conditional version of backward mean function,
$\mu_{\tau_0,\tau_1}(u)=E(V(u)|\tau_0\leq T< \tau_1)$, for $u \in
[0,\tau_0]$.  If the maximum support of $T$ is at most $\tau_1$,
then $E(V(u)|T\geq \tau_0)$ can be estimated for $u \in [0,\tau_0]$.

\begin{figure}

\includegraphics{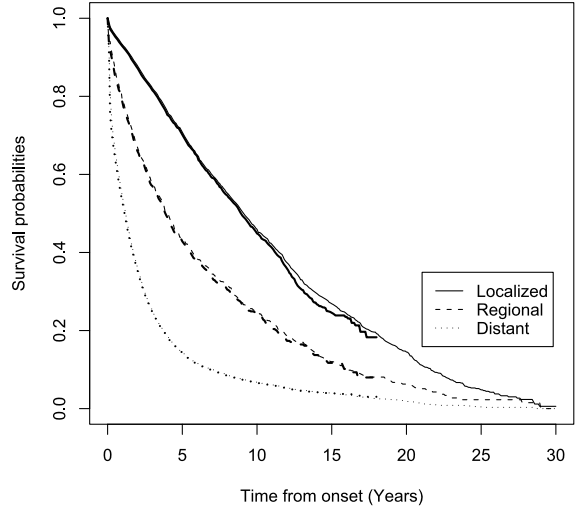}

\caption{Estimates of survival probabilities for ovarian cancer
patients in the SEER--Medicare data, using only incident cohort data
(bold) and using data from both incident and prevalent cohorts (nonbold).
Solid curves represent localized stage at diagnosis, dashed curves
represent regional stage and dotted curves represent distant stage.}\label{fig2}
\end{figure}

In an incident cohort, $\tau_1$ is usually the maximum follow-up
duration, which is determined by study design.  In a prevalent
cohort, $\tau_1$ is the longest observation time,
which is usually longer than the maximum follow-up time because subjects have already experienced the initial events before recruitment.  An
important implication of using prevalent cohort data is that it
allows us to identify a larger portion, possibly all, of the right
tail of the survival distribution.  For example, Figure \ref{fig2} shows the
estimated survival probabilities for ovarian cancer patients in
three different historic stages at diagnosis.  For all three groups
of patients, the full right tail of survival distributions can be
identified when the full data set is considered, but not in the case
when we only analyze the incident cohort.  If we only analyze the
incident cohort data, $\tau_1$ is 18 years, which is the maximum
follow-up period for the incident cohort in the data set.  When we
include prevalent cohort data in the analysis, we can estimate
$E(V(u)|T\geq \tau_0)$ nonparametrically.

\subsection{Proposed estimator}\label{sec3.3}
We propose an estimator for the backward mean function $\mu_{\tau_0,\tau_1}(u)$ by using
marked counting process arguments extending those of Huang and Louis
(\citeyear{HL1998}).  Let $N_i(t)=I(X_i\leq t,\Delta_i=1)$, $i=1,\ldots,n$, be
counting processes for observed failure, $R_i(t)=I(X_i \geq t \geq
W_i)$ be at-risk indicators,
and
\begin{eqnarray*}
N_i^V(t,u)&=& \cases{
V_i(u)I(X_i\leq t, \Delta_i=1), &\quad if $t\geq \tau_0$,\cr
0, &\quad if $ t< \tau_0$
}
\\
&=&V_i(u)I(\tau_0\leq X_i\leq t,\Delta_i=1)
\end{eqnarray*}
be marked counting processes for observed failure with a random marker
$V_i(u)$. Define averaged processes $N(t)=n^{-1}\times\sum_{i=1}^n
N_i(t)$, $N^V(t,u)=n^{-1}\times\sum_{i=1}^n N_i^V(t,u)$ and
$R(t)=n^{-1}\times\sum_{i=1}^n R_i(t)$.  Furthermore, let
$\Lambda_T(s)$ be the cumulative hazard function for $T$ and
$\Lambda^V_{\tau_0}(t,u)=\int_{\tau_0}^t E(V(u)| T=s)\Lambda_T(ds)$.
$\Lambda^V_{\tau_0}(t,u)$ can be interpreted as a hazard weighted
cumulative mean of backward processes, which is called cumulative
mark-specific hazard function in Huang and Louis (\citeyear{HL1998}).

Note that
\[
E\bigl(V(u)I(\tau_0\leq T<\tau_1)\bigr)=\int_{\tau_0}^{\tau_1}
S(s) \Lambda^V_{\tau_0}(ds,u).
\]
If we have an estimate of $\Lambda^V_{\tau_0}(t,u)$, denoted by
$\hat{\Lambda}^V_{\tau_0}(t,u)$, then we can estimate $E(V(u)I(\tau_0\leq
T<\tau_1))$ by $\int_{\tau_0}^{\tau_1}
\hat{S}(s) \hat{\Lambda}^V_{\tau_0}(ds,u)$,\vspace*{1pt} where $\hat{S}(t)$ is the
product limit estimate using left truncated and right censored data [Tsai, Jewell and Wang
(\citeyear{TJW1987}), Lai and Ying (\citeyear{LY1991})]. Since
\begin{eqnarray*}
\Lambda_{\tau_0}^V(t,u)&=&\int_{\tau_0}^tE\bigl(V(u)|T=s\bigr)\Lambda_T(ds)\\
&=&\int_{\tau_0}^t\frac{E(V(u)I(T=s))P(C\geq s \geq
W)/\beta}{S(s)P(C\geq s \geq W)/\beta} \,ds=\int_{\tau_0}^t
\frac{E(N^V(ds,u))}{G(s)},
\end{eqnarray*}
where the expectation is taken conditioning on $T \geq W$,
$\Lambda^V_{\tau_0}(t,u)$ can be estimated by
\begin{equation}\label{E:GNA2}
\hat{\Lambda}^V_{\tau_0}(t,u)=\int_{\tau_0}^t \frac{N^V(ds,u)}{R(s)}.
\end{equation}
The backward mean function $\mu_{\tau_0,\tau_1}(u)$ can then be estimated by

\begin{eqnarray}\label{E:est2}
\hat{\mu}_{\tau_0,\tau_1}(u)&=&\frac{1}{\hat{S}(\tau_0)-\hat{S}(\tau_1)}\int_{\tau_0}^{\tau_1} \hat{S}(s)
\hat{\Lambda}^V_{\tau_0}(ds,u)\nonumber\\[-8pt]\\[-8pt]
&=&\frac{1}{n}\frac{1}{\hat{S}(\tau_0)-\hat{S}(\tau_1)}\sum_{i=1}^n
\frac{\hat{S}(X_i)\Delta_i V_i(u)I(\tau_0\leq X_i<\tau_1)}{{R}(X_i)}.\nonumber
\end{eqnarray}

More generally, we can estimate $\mu_{t_1,t_2}(u)=E(V(u)|t_1\leq T <
t_2)$ for $u\leq t_1<t_2\leq \tau_1$ and $u \in [0,\tau_0]$,
which can be estimated by
\[
\hat{\mu}_{t_1,t_2}(u)=\frac{1}{n}\frac{1}{\hat{S}(t_1)-\hat{S}(t_2)}\sum_{i=1}^n
\frac{\hat{S}(X_i)\Delta_i V_i(u)I(t_1\leq X_i<t_2)}{R(X_i)}.
\]

The mean of $V(u)$ can be estimated as long as $T>u$.  However, if we
estimate $E(V(u)|u\leq T<\tau_1)$, the subpopulation defined by
conditioning changes with the time index $u$, and the estimand loses a
desirable interpretation of being a mean process for a fixed underlying
population.  Although the introduction of the constant $\tau_0$ in the
conditioning may not use information from part of the data, it defines
a meaningful subpopulation such that the whole backward function can be
studied for the same underlying population.

The following theorem states the large sample properties of the
proposed estimator.

\begin{theorem}\label{thm3.1}
Assume that $E(V^2(\tau_0))<\infty$ and certain technical restrictions
on the support of $(T,C,W)$ hold [Wang (\citeyear{W1991})].  For $\tau_0\leq
t_1<t_2\leq \tau_1$, $\hat{\mu}_{t_1,t_2}(u) \to {\mu}_{t_1,t_2}(u)$
uniformly a.s. on $[0,\tau_0]$. Also,\vspace*{1pt}
$n^{1/2}(\hat{\mu}_{t_1,t_2}(u)-{\mu}_{t_1,t_2}(u))=n^{-1/2}\sum_{i=1}^n
\xi_i(u)+o_p(1)$, where $\xi_i(u)$ is defined in the \hyperref[app]{Appendix} and the
random sequence converges weakly to a Gaussian process with covariance
function
\begin{eqnarray}\label{E:cst}
C_{t_1,t_2}(u,v)&=&\frac{1}{(S(t_1)-S(t_2))^2}\int_{t_1}^{t_2}\frac{S^2(s)}{G(s)}E\bigl(V(u)V(v)|T=s\bigr)\Lambda_T(ds)\nonumber\\
&&{}-\frac{1}{(S(t_1)-S(t_2))^3}\int_{t_1}^{t_2}\frac{S(s)H_{t_1,t_2}(s,v)}{G(s)}\Lambda^V(ds,u)\nonumber\\[-8pt]\\[-8pt]
&&{}-\frac{1}{(S(t_1)-S(t_2))^3}\int_{t_1}^{t_2}\frac{S(s)H_{t_1,t_2}(s,u)}{G(s)}\Lambda^V(ds,v)\nonumber\\
&&{}+\frac{1}{(S(t_1)-S(t_2))^4}\int_{t_1}^{t_2}\frac{H_{t_1,t_2}(s,u)H_{t_1,t_2}(s,v)}{G(s)}\Lambda_T(ds),\nonumber
\end{eqnarray}
where
\[
H_{t_1,t_2}(s,u)=E\bigl(V(u)I(s\leq T<t_2)\bigr)S(t_1) +E\bigl(V(u)I(t_1\leq
T<s)\bigr)S(t_2).
\]
\end{theorem}

From (\ref{E:cst}), $C_{t_1,t_2}(u,v)$ can be consistently estimated
by
\begin{eqnarray}\label{E:sigmast}
\hat{\Sigma}_{t_1,t_2}(u,v)&=&\frac{1}{n}\sum_{i=1}^n\frac{\Delta_iI(t_1\leq
X_i<t_2)}{R^2(X_i)(\hat{S}(t_1)-\hat{S}(t_2))^2}
 \biggl[\hat{S}(X_i)V_i(u)-\frac{\hat{H}_{t_1,t_2}(X_i,u)}{(\hat{S}(t_1)-\hat{S}(t_2))}
 \biggr]\nonumber\\[-8pt]\\[-8pt]
&&\hphantom{\frac{1}{n}\sum_{i=1}^n}{}\times \biggl[\hat{S}(X_i)V_i(v)-\frac{\hat{H}_{t_1,t_2}(X_i,v)}{(\hat{S}(t_1)-\hat{S}(t_2))} \biggr],\nonumber
\end{eqnarray}
where
\begin{eqnarray*}
\hat{H}_{t_1,t_2}(s,u)&=&\frac{1}{n}\sum_{j=1}^n\bigl(\Delta_j
V_j(u)\hat{S}(X_j)[I(t_1\leq s\leq X_j<t_2)\hat{S}(t_1)\\
&&\hphantom{\frac{1}{n}\sum_{j=1}^n\bigl(\Delta_j V_j(u)\hat{S}(X_j)[}
{}+I(t_1\leq
X_j<s\leq t_2)\hat{S}(t_2)]\bigr)/R(X_j).
\end{eqnarray*}

\subsection{Construction of confidence bands}\label{sec3.4}
From the large sample results, we can construct pointwise confidence
intervals in the form $\hat{\mu}_{t_1,t_2}(u)\pm\break
n^{-1/2}z\hat{\sigma}_{t_1,t_2}(u)$, where $z$ is the standard normal
critical value and
$\hat{\sigma}_{t_1,t_2}(u)=\hat{\Sigma}^{1/2}_{t_1,t_2}(u,u)$. Since we
are estimating mean functions of processes, it is also of interest to
construct confidence bands for a given level of significance.   We will
replace $z$ in a pointwise confidence interval by a larger value $b$ to
reach an appropriate simultaneous coverage probability. Although
$\sqrt{n}(\hat{\mu}_{t_1,t_2}(u)-{\mu}_{t_1,t_2}(u))$ converges to a
Gaussian process, the limiting process may not have independent
increment because $V(u)$ is an arbitrary process.  Thus, it may not be
possible to compute the exact asymptotic distribution. To construct
confidence bands, we approximate the limiting process by a multiplier
bootstrap method described as follows:

\begin{enumerate}
\item Generate random multipliers $\{G_i, i=1,\ldots,n\}$
which are independent standard normal distributed and independent of
the data. Then, compute
\[
W(u)=n^{-1/2}\sum_{i=1}^n G_i \biggl\{\frac{\Delta_iI(t_1\leq
X_i<t_2)}{R(X_i)(\hat{S}(t_1)-\hat{S}(t_2))}
 \biggl[\hat{S}(X_i)V_i(u)-\frac{\hat{H}_{t_1,t_2}(X_i,u)}{(\hat{S}(t_1)-\hat{S}(t_2))} \biggr] \biggr\}.
\]

\item Repeat step 1 until $m$ versions of $W(u)$ are
obtained, denoted by $\{W_k(u),k=1,\ldots,m\}$.

\item Obtain $b$ which is the
$(100-\alpha)$-percentile of
$\max_{(0,\tau_0)}\{|W_k(u)|\}$.

\item The confidence band for $\mu_{t_1,t_2}(u), u
\in [0,\tau_0]$, can be calculated by $\hat{\mu}_{t_1,t_2}(u)\pm n^{-1/2}b\hat{\sigma}_{t_1,t_2}(u).$
\end{enumerate}

The above method uses the simulated samples $W(u)$ to approximate the
distribution of $\xi(u)$, the influence function of
$\hat{\mu}_{t_1,t_2}(u)$. The method is motivated by the construction
of confidence bands for survival function in a proportional hazards
model proposed by Lin, Fleming and Wei (\citeyear{LFW1994}). By the
permanence of the Donsker property [van der Vaart and Wellner (\citeyear{VW1996})],
$W(u)$ can be shown to converge to a Gaussian process. Also,
conditional on observed data, $E(W(u)W(v))$ equals the right-hand side
of (\ref{E:sigmast}) since $E(G_iG_j)=0$ for $i \neq j$ and
$E(G_i^2)=1$. Hence, $E(W(u)W(v))$ converges almost surely to
$C_{t_1,t_2}(u,v)$ and $W(u)$ has the same asymptotic distribution as
$\sqrt{n}(\hat{\mu}_{t_1,t_2}(u)-\mu_{t_1,t_2}(u))$.

In the medical cost example, $\mu_{t_1,t_2}(u)$ is nonnegative and it
is more meaningful to construct confidence intervals or confidence
bands that are always nonnegative. For this purpose, we consider the
log-transformation and the confidence bands have the form
$\hat{\mu}_{t_1,t_2}(u)\exp[\pm
n^{-1/2}b^*\hat{\sigma}_{t_1,t_2}(u)/\hat{\mu}_{t_1,t_2}(u)]$.  $b^*$
can be found by the above algorithm with a slight modification in step
3, where $b^*$ is the $(100-\alpha)$-percentile of
$\max_{(0,\tau_0)}[|W_k(u)|/\hat{\sigma}_{t_1,t_2}(u)]$. The reason is
that by the functional delta method,
$\sqrt{n}\mu_{t_1,t_2}(u)(\ln{\hat{\mu}_{t_1,t_2}(u)}-\ln{\mu_{t_1,t_2}(u)})/\sigma_{t_1,t_2}(u)$
is asymptotically equivalent to
$(\mu_{t_1,t_2}(u)/{\sigma_{t_1,t_2}(u)})\times({1}/{\mu_{t_1,t_2}(u)})\times
\xi(u)={\xi(u)}/{\sigma_{t_1,t_2}(u)}$, whose distribution is
approximated by $W(u)/\hat{\sigma}_{t_1,t_2}(u)$.

\section{Extensions of estimation}\label{sec4}
\subsection{Distribution and percentile estimation}\label{sec4.1}

In a lot of applications, including medical cost, the distribution of $V(u)$ is not symmetric and is
often highly skewed. In these cases, apart from estimating the mean function, one
might also be interested in estimating percentile and distribution
functions of $V(u)$. Here we extend the marked process approach of
mean estimation to estimate a joint distribution function,
$p_{\tau_0,\tau_1}(m,t,u)=P(V(u)\leq m,T\leq t|\tau_0\leq T<\tau_1)$
where $\tau_0\leq t<\tau_1$.  Let
\begin{eqnarray*}
\tilde{N}_i^V(t,u)&=& \cases{
I\bigl(V_i(u)\leq m, X_i\leq t, \Delta_i=1\bigr), &\quad if $ t\geq \tau_0$,\cr
0, &\quad if $t< \tau_0$
}
 \\
 &=&I\bigl(V_i(u)\leq m,\tau_0\leq X_i\leq t,\Delta_i=1\bigr)
\end{eqnarray*}
and $\tilde{\Lambda}^V_{\tau_0}(t,u)=\int_{\tau_0}^tP(V(u)\leq m
|T=s) \Lambda_T(ds)$.  Following
the arguments in Section \ref{sec3}, $p_{\tau_0,\tau_1}(m,t,u)$ can be
estimated by
\[
\hat{p}_{\tau_0,\tau_1}(m,t,u)=\frac{1}{n}\frac{1}{\hat{S}(\tau_0)-\hat{S}(\tau_1)}
\sum_{i=1}^n\frac{\hat{S}(X_i)\Delta_iI(V_i(u)\leq m, \tau_0\leq
X_i\leq t)}{R(X_i)}.
\]
This joint distribution estimate can be used for correlation
analysis between $V(u)$ and $T$.  Similar to the estimator of mean
function, $\hat{p}_{\tau_0,\tau_1}(m,t,u)$ is a consistent estimate
of $p_{\tau_0,\tau_1}(m,t,u)$.

Next, we consider estimating a pointwise $q$th-percentile
function, $m^q_{\tau_0,\tau_1}(u)$, which is defined by
\[
P\bigl(V(u)\leq m^q_{\tau_0,\tau_1}(u)|\tau_0\leq T<\tau_1\bigr)=q
\]
for $u\in[0,\tau_0]$ and $0<q<1$. To estimate
$m^q_{\tau_0,\tau_1}(u)$, consider the estimating function
\begin{eqnarray}\label{E:quantileesteq}
\varphi_q(m,u)&=&\frac{1}{n}\frac{1}{\hat{S}(\tau_0)-\hat{S}(\tau_1)}\nonumber\\[-8pt]\\[-8pt]
&&{}\times\sum_{i=1}^n
\frac{\hat{S}(X_i)\Delta_iI(\tau_0\leq X_i<\tau_1)(I(V_i(u)\leq
m)-q)}{R(X_i)}.\nonumber
\end{eqnarray}
It can be seen that $\varphi_q(m_0,u)$ converges in probability to
$0$ for $m_0=m^q_{\tau_0,\tau_1}(u)$.  Thus, a natural estimator of
$m^q_{\tau_0,\tau_1}(u)$ is the zero-crossing of $\varphi_q(m,u)$.
The existence of a solution is guaranteed because $\varphi_q(m,u)$
is increasing in $m$ and $\lim_{m\to-\infty}\varphi_q(m,u)<0$ and
$\lim_{m\to \infty}\varphi_q(m,u)>0$, for $0<q<1$.  The estimation
of $m^q_{\tau_0,\tau_1}(u)$ can be easily implemented in common
statistical softwares by noting from (\ref{E:quantileesteq}) that
this quantity can be estimated by a weighted empirical percentile of
$V(u)$ with weights equals to $\hat{S}(X_i)\Delta_iI(\tau_0\leq
X_i<\tau_1)/R(X_i)$.

\subsection{Estimation of backward rate function}\label{sec4.2}
When the mean rate of change of stochastic processes before failure events is of scientific interest, one
might want to estimate an associated quantity
$r(u)=E(\frac{dV(u)}{du})$.  In the medical cost example, $r(u)$ is the mean rate of cost
accrual \emph{per unit time} at $u$ time units before a failure event. $r(u)$ is a
measure of instantaneous change in the backward stochastic process. We call
$r(u)$ the backward rate function.

Like the estimation of backward mean functions, we can only estimate nonparametrically a conditional
version $r_{\tau_0,\tau_1}(u)=E(\frac{dV(u)}{du}|\tau_0\leq
T<\tau_1)$. Similar to $\mu_{\tau_0,\tau_1}(u)$, we have the
following relationship:
\begin{eqnarray}\label{E:r}
r_{\tau_0,\tau_1}(u)\bigl(S(\tau_1)-S(\tau_0)\bigr)&=&E \biggl(\frac{dV(u)}{du}I(\tau_0\leq
T<\tau_1) \biggr)\nonumber\\[-8pt]\\[-8pt]
&=&\int_{\tau_0}^{\tau_1}S(s)E\biggl(\frac{dV(u)}{du}\Big|T=s\biggr)\Lambda_T(ds).\nonumber
\end{eqnarray}
To estimate $r_{\tau_0,\tau_1}(u)$, it suffices to estimate
$E(\frac{dV(u)}{du}|T=s)$ at the jump points of $\hat{\Lambda}_T$,
which are the uncensored survival times.  For each uncensored\vspace*{1pt}
individual, $E(\frac{dV(u)}{du}|T_i)$ can be estimated by
\[
\hat{v}_i(u)=\frac{1}{h}\int_0^{\tau_0}
k \biggl(\frac{u-v}{h} \biggr) \,dV_i(v),
\]
where $k(\cdot)$ is a kernel function satisfying $\int_0^{\tau_0}
k(s) \,ds=1$ and $h>0$ is a bandwidth parameter, which can be chosen by minimizing an integrated mean square error.  $\hat{v}_i(u)$ is
similar in nature to the estimator of the subject specific rate of
recurrent event proposed by Wang and Chiang (\citeyear{WC2002}). Substituting unknown
quantities in (\ref{E:r}) by their estimates, $r_{\tau_0,\tau_1}(u)$
can be estimated by
\begin{eqnarray}\label{E:estr}
\hat{r}_{\tau_0,\tau_1}(u)=\frac{1}{n}\frac{1}{\hat{S}(\tau_0)-\hat{S}(\tau_1)}\sum_{i=1}^n
\frac{S(X_i)\Delta_i \hat{v}_i(u)I(\tau_0\leq X_i<\tau_1)}{R(X_i)}.
\end{eqnarray}
The estimator (\ref{E:estr}) can also be derived as the convolution
smoothing estimator of $\hat{\mu}_{\tau_0,\tau_1}$.  It can be shown
that
\[
\hat{r}_{\tau_0,\tau_1}(u)=\frac{1}{h}\int_0^{\tau_0}
k \biggl(\frac{u-v}{h} \biggr) \hat{\mu}_{\tau_0,\tau_1}(dv).
\]

\section{Numerical studies}\label{sec5}
\subsection{Simulations}
Finite sample performance of the proposed estimator in Section \ref{sec3} and
the empirical coverage of pointwise confidence intervals and overall confidence bands are evaluated by
simulations. Data are generated 2000 times in each simulation, and
each simulated data set consists of 100 or 400 observations. The
confidence bands are constructed by simulating 1000 sets of random multipliers in
each simulated data set.

The simulation follows data structure similar to the SEER--Medicare
linked data.  We generated survival time $T$ from a gamma distribution
with shape 3 and rate 1, truncation time $W$ has half chance to be $0$
and half chance to be generated from a uniform-$(0,20)$ distribution,
and censoring time $C=W+C'$ where $C'$ is generated from a
uniform-$(0,8)$ distribution. The subset with truncation time $W=0$
represents an incident cohort and $W>0$ a prevalent cohort with
untruncated observations satisfying $T\geq W$. Conditioning on $T$, we
generated two independent latent variables $Z_1$ and $Z_2$ from a gamma
distribution with shape 3 and rate $T$. The latent variables are used
to induce correlation between survival time and stochastic processes.
For each subject, we generate a recurrent event process $P(\cdot)$ from
a Poisson process with rate $4Z_1$, and at each occurrence of recurrent
events at $u$ time units before failure event, a variable  $Q(u)$ is
generated from a gamma distribution with shape $Z_2\times[3+3\times
I(u<1/3)]$ and rate 1. The process of interest is $V(u)=\int_{T-u}^T
Q(s)\,dP(s)$.  The generated data has the same structure as medical cost
data, where $P$ represents counting process for recurrent
hospitalizations, $Q$ represents medical cost incurred at a particular
hospitalization and $V(u)$ is the total medical cost in the last $u$
time units of life.  The recurrent event process, medical cost process
and failure time are correlated through latent variables.   That is,
medical cost processes are terminated by informative failure events.
Our simulations generated negative correlation between end-of-life cost
and failure time, which also matches with the SEER--Medicare linked data
(see Section~\ref{sec5.2}).  Under this setting, we are interested in estimating
$E(V(u)|1\leq T<20)$ for $u\in [0,1]$.\looseness=1

We compare the proposed estimator with naive complete-case estimators
that have been used in the medical literature. Supposing one uses an
unweighted sample mean based on observed deaths for the analysis, the
direction of bias for this naive analysis will depend on whether longer
survivors or shorter survivors are being oversampled.  In an incident
cohort, naive analysis will oversample shorter survivors in general
because the naive data set is right truncated by discarding the right
censored observations. So the estimated mean end-of-life cost will be
biased upward in the simulation.  In a prevalent cohort, naive sample
is subject to double truncation, but the effect from left truncation is
more serious in the simulation and we oversample longer survivors in
general, so the estimated mean end-of-life cost will be biased
downward.  The simulation results are shown in Table~\ref{table1} and match with
this reasoning.  The proposed method can correct the bias caused by
left truncation and right censoring.  The unweighted complete case
estimator has been used, for example, in Chan et al. (\citeyear{Cetal1995}), for
studying the frequency of opportunistic infections for HIV infected
individuals before death.\looseness=1

The small sample bias of the proposed estimator and evaluation of the variance estimator is also shown in Table \ref{table1}.  We can
see that the proposed estimator worked well in practical sample
sizes.  We also studied the empirical coverage of the $95\%$
confidence bands.  Let $t^*=\min\{u\dvtx V_i(u)>0$ for some $i\}$.  Since
$\mu_{t_1,t_2}(u)=0$ for $u< t^*$, it is only meaningful to consider
coverage probabilities for $u\geq t^*$.  We considered the coverage of confidence band for $u \in [t^*,1]$.
The empirical coverage of the 95\% confidence bands are $94\%$ for
both $n=100$ and $n=400$.  The empirical coverage of the confidence bands are close to the
nominal value for practical sample sizes.

\begin{table}
\caption{Summary of the simulation study:  Comparisons among the
proposed estimator and naive estimators based on unweighted complete
case analysis from incident and prevalent cohorts, and evaluation of
variance estimates and pointwise coverage probabilities of 95\%
confidence intervals using the proposed methodologies.  SSE represents
the sampling standard deviation and SEE is the~sample~average of the
standard error estimates}\label{table1}
\begin{tabular*}{\textwidth}{@{\extracolsep{\fill}}ld{1.1}d{2.2}d{2.2}d{2.2}d{2.2}ccc@{}}
\hline
\multirow{2}{29pt}[-6pt]{\textbf{Sample size}}&&&\multicolumn{2}{c}{\textbf{Naive estimators}}&\multicolumn{4}{c@{}}{\textbf{Proposed estimators}}\\[-5pt]
&&&\multicolumn{2}{c}{\hrulefill}&\multicolumn{4}{c@{}}{\hrulefill}\\
&\multicolumn{1}{c}{$\bolds{u}$}&\multicolumn{1}{c}{\textbf{Truth}}&\multicolumn{1}{c}{\textbf{Incident}}&\multicolumn{1}{c}{\textbf{Prevalent}}&
\multicolumn{1}{c}{\textbf{Estimate}}&\textbf{SSE}&\textbf{SEE}&\textbf{Coverage}\\
\hline
100&0.1&  4.32&   5.34&   2.27&   4.19&   1.24&   1.35&   0.92\\
&0.2&   8.64&   10.71&  4.53&   8.41&   2.13&   2.25&   0.92\\
&0.3&   12.96&  16.08&  6.79&   12.62&  3.00&   3.13&   0.93\\
&0.4&   15.84&  19.61&  8.38&   15.4&   3.54&   3.66&   0.93\\
&0.5&   18.00&  22.28&  9.53&   17.5&   3.94&   4.01&   0.93\\
&0.6&   20.16&  24.92&  10.64&  19.57&  4.35&   4.40&   0.93\\
&0.7&   22.32&  27.52&  11.8&   21.62&  4.73&   4.78&   0.93\\
&0.8&   24.48&  30.08&  12.92&  23.64&  5.11&   5.12&   0.93\\
&0.9&   26.64&  32.61&  14.01&  25.63&  5.46&   5.45&   0.93\\
&1.0&   28.80&  35.08&  15.11&  27.58&  5.79&   5.77&   0.93\\[6pt]
400& 0.1& 4.32&   5.39&   2.22&   4.29&   0.67&   0.69&   0.94\\
&0.2&   8.64&   10.78&  4.38&   8.58&   1.13&   1.14&   0.95\\
&0.3&   12.96&  16.17&  6.60&   12.86&  1.58&   1.60&   0.95\\
&0.4&   15.84&  19.75&  8.07&   15.72&  1.87&   1.88&   0.95\\
&0.5&   18.00&  22.44&  9.19&   17.86&  2.08&   2.07&   0.95\\
&0.6&   20.16&  25.11&  10.29&  19.98&  2.29&   2.26&   0.95\\
&0.7&   22.32&  27.76&  11.41&  22.09&  2.49&   2.44&   0.95\\
&0.8&   24.48&  30.35&  12.52&  24.17&  2.68&   2.62&   0.95\\
&0.9&   26.64&  32.89&  13.61&  26.20&  2.86&   2.77&   0.95\\
&1.0&   28.80&  35.36&  14.69&  28.18&  3.03&   2.93&   0.95\\
\hline
\end{tabular*}
\end{table}

\subsection{Data analysis}\label{sec5.2}
The proposed methods are illustrated by analyzing the SEER--Medicare
linked data.  We investigated end-of-life-cost for ovarian cancer
cases diagnosed at age 65 or older among Medicare enrolles. Total
amount charged during hospitalization is considered as medical cost
in the analysis; this includes charges not covered by Medicare. All
medical expenditures are adjusted to January 2000 value by the
medical care component of the consumer price index, available from
the website of the U.S. Department of Labor (\url{http://www.bls.gov/cpi/}).
We compare medical cost among individuals with different historic
stages at diagnosis.  There were 3766, 1400 and 15,104 subjects
classified as localized, regional and distant stages at diagnosis
respectively. The estimates of the survival probabilities are shown
in Figure \ref{fig2}.

\begin{figure}

\includegraphics{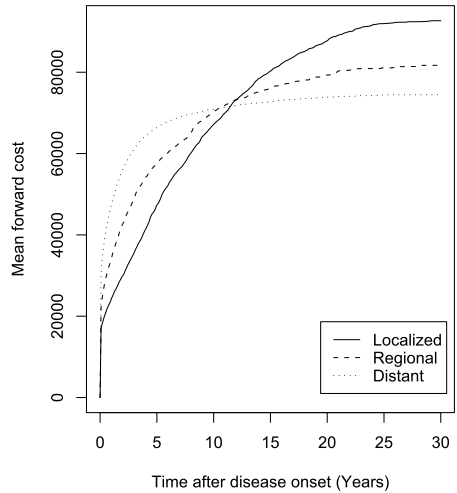}

\caption{Estimates of the mean forward cost functions for ovarian
cancer patients.  Solid curve represents localized stage at
diagnosis, dashed curve represents regional stage and dotted curve
represents distant stage.}\label{fig3}
\end{figure}

First, we compare the estimated mean forward cost functions among the
three historic stages.  A mean forward cost process is estimated
by
\[
\hat{\mu}_Y(t)=\frac{1}{n}\sum_{i=1}^n \int_0^t
\frac{\hat{S}(s)I(W_i\leq s \leq C_i)\, dY_i(s)}{R(s)},
\]
which can be viewed as a limiting case of the estimator of Lin et al.
(\citeyear{Letal1997}) with the partition size tending to zero. If left truncation is
absent, methods proposed by Bang and Tsiatis (\citeyear{BT2000}), Strawderman (\citeyear{S2000})
and Zhao and Tian (\citeyear{ZT2001}) can also be extended to estimate forward mean
functions.  Figure \ref{fig3} shows the estimates for mean forward cost
functions up to the thirtieth year after initial diagnosis of cancer.
Note that there is a crossover for three curves around ten years after
diagnosis.  The ten year estimated survival probabilities are 0.47,
0.25, 0.07 (s.e.: 0.03, 0.06, 0.05) for patients diagnosed with local,
regional and distinct stages respectively.  In the first ten years
after diagnosis, cumulative cost reflects the severity of the cancer
stage at diagnosis.  Beyond the tenth year, the cumulative cost
reflects the better chance of survival for the less severe stages of
cancer. The conflicting nature between accumulation of cost and
survival complicates the analysis and careful interpretation of the
results are needed. Also, the forward cost functions cannot directly
answer questions about end-of-life-cost, because individuals have
different survival times and the increase in medical cost before
failure events at a given time after disease onset is offset by
nonsurvivors who do not contribute to any increase in medical cost
after death.

\begin{figure}

\includegraphics{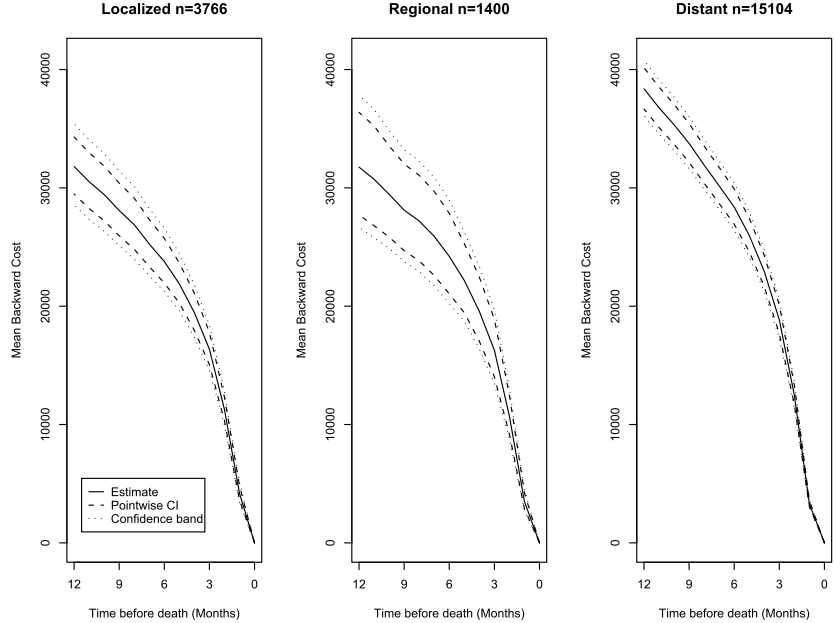}

\caption{Estimates of the mean backward cost functions for ovarian
cancer patients.  Solid curves represent the estimates.  Dotted curves
represent 95\% simultaneous confidence bands.  Dashed curves represent
pointwise 95\% confidence intervals.}\label{fig4}
\end{figure}

\begin{figure}

\includegraphics{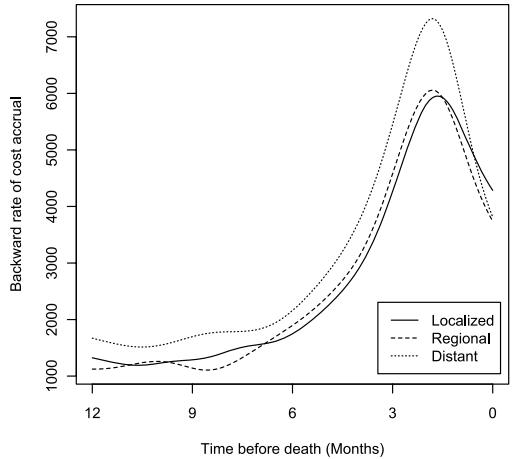}

\caption{Estimates of the backward rate of cost accrual.
Solid curve represents localized stage at diagnosis, dashed curve
represents regional stage and dotted curve represents distant stage.}\label{fig5}
\end{figure}

In the SEER--Medicare data analysis we observe a negative correlation
between end-of-life-cost and survival.  Using the estimator of joint
distribution in Section~\ref{sec4.1}, the estimated Pearson correlation
coefficient between $V(1)$ and $T$ (conditioned on $T> \tau_0=1$) is
$-0.65$, $-0.31$ and $-0.46$ for localized, regional and distant stages
respectively.  We compare the estimated one-year mean backward cost
functions among the three historic stages, for individuals surviving at
least one year after onset of disease.  The results are shown in Figure
\ref{fig4}. Unlike mean forward cost functions, estimated backward cost
functions are very similar in shape for the three historic stage
groups. The results show that there is a terminal increase in medical
cost before death.  The estimated final-year medical cost of a patient
is \$31,802, \$31,752, \$38,377 (s.e.: \$1229, \$2205, \$896) in January
2000 value for patients diagnosed with local, regional and distinct
stages respectively. The estimated medical cost for the last three
months of life of an ovarian cancer patient is \$16,365, \$16,284,
\$18,848 (s.e.: \$692, \$1236, \$613) in January 2000 value for patients
diagnosed with local, regional and distinct stages respectively. Figure
\ref{fig5} shows the backward rate of cost accrual, which is the end-of-life
cost per unit time before death.  The bandwidths for the estimates were
chosen to minimize an integrated mean squared error.  The results agree
with Figure \ref{fig4} that there is a terminal increase in medical cost before
death.

\section{Concluding remarks}\label{sec6}
In this paper we proposed statistical methods for studying the terminal
behavior of stochastic processes before failure events. In particular,
we discussed nonparametric methods for estimating the mean function of
backward stochastic processes under incident and prevalent sampling
designs.  We also discussed identifiability issues related to
estimation with incomplete follow-up data.  In an incident sampling
design, the right tail of survival distribution may not be identified
because of limited study duration. Using prevalent sampling design, the
identifiable region for the survival distribution could be enlarged and
cost associates with those individuals can be identified.

We used the SEER--Medicare data as an example throughout this paper.
Although the SEER--Medicare data contain both incident and prevalent
cohorts, our method can be applied to data only from an incident cohort
or a prevalent cohort.  The proposed methods only require the
stochastic process data to be available in a certain time interval
before a failure event. Thus, prevalent data can be used alone for the
proposed methods even though we do not have data on the stochastic
process before patient enrollment.

The backward estimation procedure proposed in this paper could serve as
a main building block for other analyses.  For example, we can compute
the ratio of end-of-life cost to lifetime cost that combine the
proposed method and the existing methods for analyzing lifetime cost.

Although we use medical cost as an example, applications of the
proposed methods do not only limit one to study medical cost, but can also be
used to study the terminal behavior of other stochastic processes
before failure events. Other applications include CD4 counts
before AIDS diagnosis, frequency of hospitalizations before death and
measurements of quality-of-life before death.

The main focus of this paper is one sample estimation of the
backward mean function.  The authors are extending the idea in this
paper to regression models of backward mean functions and backward
rate functions.

\begin{appendix}
\section*{\texorpdfstring{Appendix: Proof of Theorem \protect\lowercase{\ref{thm3.1}}}{Appendix: Proof of Theorem 3.1}}\label{app}

We apply empirical process theory to
prove the asymptotic results. Since $N^V(t,u)$ is having bounded variations and $E(N^V(t,u))<\infty$ for
$(t,u)\in[\tau_0,\tau_1]\times[0,\tau_0]$,\vspace*{1pt} we can apply the uniform
strong law of large numbers [Pollard (\citeyear{P1990})] to show that $N^V(t,u)$
converges a.s. uniformly to $E(N^V(s,u))=\int_{\tau_0}^s
E(V(u)I(T=s))P(C\geq s\geq W)/\beta  \,ds$.\vspace*{1pt} Also, $R(s)$ converges
a.s. uniformly to $G(s)$ [Woodroofe (\citeyear{W1985})].  By Lemma 1 of
Lin et al. (\citeyear{Letal2000}),
\[
\hat{\Lambda}^V_{\tau_0}(t,u)=\int_{\tau_0}^t
\frac{N(ds,u)}{R(s)}\stackrel{\mathrm{a.s.}}{\rightarrow} \int_{\tau_0}^t
\frac{E(N(ds,u))}{G(s)}=\Lambda^V_{\tau_0}(t,u)
\]
uniformly on $[\tau_0,\tau_1]\times[0,\tau_0]$.  Also, since
$\hat{S}(t)$ and $\hat{\Lambda}^V_{\tau_0}(t,u)$ are uniform
consistent estimates of $S(t)$ and $\Lambda^V_{\tau_0}(t,u)$, uniform\vspace*{1pt}
consistency of $\hat{\mu}_{t_1,t_2}(u)$ also follows from Lemma 1 in Lin
et al. (\citeyear{Letal2000}).

Defining $M_i(t)=N_i(t)-\int_0^t R_i(s)\Lambda_T(ds)$ for $t\geq0$,
$M_i^V(t,u)=N_i^V(t,u)-\int_{\tau_0}^t R_i(s)\Lambda_{\tau_0}^V(ds,u)$
for $t\geq \tau_0$ and $u\in[0,\tau_0]$, we have
\begin{eqnarray*}
&&\sqrt{n}\bigl(\hat{\Lambda}^V_{\tau_0}(t,u)-{\Lambda}^V_{\tau_0}(t,u)\bigr)\\
&&\qquad=\sqrt{n} \biggl\{\int_{\tau_0}^t\frac{N(ds,u)}{R(s)}-\int_{\tau_0}^t\frac{E(N(ds,u))}{G(s)} \biggr\}\\
&&\qquad=\int_{\tau_0}^t\frac{\sqrt{n}(N(ds,u)-E(N(ds,u)))}{G(s)}\\
&&\quad\qquad{}-\int_{\tau_0}^t\frac{\sqrt{n}(R(s)-G(s))}{G^2(s)}E(N(ds,u))+o_p(1)\\
&&\qquad=\int_{\tau_0}^t
\frac{\sqrt{n}N(ds,u)}{G(s)}-\int_{\tau_0}^t\frac{\sqrt{n}R(s)}{G(s)}\Lambda^V_{\tau_0}(ds,u)+o_p(1)\\
&&\qquad=n^{-1/2}\sum_{i=1}^n\int_{\tau_0}^t\frac{M_i^V(ds,u)}{G(s)}+o_p(1)
\end{eqnarray*}
and
\begin{eqnarray*}
&&\sqrt{n}\bigl(\hat{\mu}_{t_1,t_2}(u)-{\mu}_{t_1,t_2}(u)\bigr)\\
&&\qquad=\sqrt{n}
\biggl(\frac{1}{\hat{S(t_1)}-\hat{S}(t_2)}\int_{t_1}^{t_2}\frac{\hat{S}(s)N^V(ds,u)}{R(s)}\\
&&\quad\qquad\hphantom{\sqrt{n}\biggl(}
{}-\frac{1}{{S(t_1)}-{S}(t_2)}\int_{t_1}^{t_2}\frac{{S}(s)E(N^V(ds,u))}{G(s)} \biggr)\\
&&\qquad=-\frac{\sqrt{n}[(\hat{S}(t_1)-S(t_1))-(\hat{S}(t_2)-S(t_2))]}{(S(t_1)-S(t_2))^2}\int_{t_1}^{t_2}S(t)\Lambda^V_{\tau_0}(ds,u)\\
&&\quad\qquad{}+\frac{1}{{S(t_1)}-{S}(t_2)}\int_{t_1}^{t_2}S(s)\sqrt{n}\bigl(\hat{\Lambda}^V_{\tau_0}(ds,u)-\Lambda^V_{t_1}(ds,u)\bigr)\\
&&\quad\qquad{}+\frac{1}{{S(t_1)}-{S}(t_2)}\int_{t_1}^{t_2}\int_{t_1}^{t_2}\sqrt{n}\bigl(\hat{S}(s)-S(s)\bigr)\Lambda^V_{\tau_0}(ds,u)+o_p(1)\\
&&\qquad=n^{-1/2}\sum_{i=1}^n \bigl(\xi_{1i}(u)+\xi_{2i}(u)+\xi_{3i}(u)\bigr)+o_p(1),
\end{eqnarray*}
where
\begin{eqnarray*}
\xi_{1i}(u)&=&\frac{E(V(u)|t_1\leq T<t_2)}{S(t_1)-S(t_2)} \biggl[S(t_1)\int_0^{t_1}\frac{M_i(dt)}{G(s)}-S(t_2)\int_0^{t_2}\frac{M_i(dt)}{G(s)} \biggr],\\
\xi_{2i}(u)&=&\frac{1}{S(t_1)-S(t_2)}\int_{t_1}^{t_2}\frac{S(s)M_i^V(ds,u)}{G(s)},\\
\xi_{3i}(u)&=&-\frac{1}{S(t_1)-S(t_2)}\int_{t_1}^{t_2}S(s)\int_0^s
\frac{M_i(dt)}{G(s)} \Lambda^V(ds,u).
\end{eqnarray*}
Upon algebraic manipulation, $\xi_i(u)=\xi_{1i}(u)+\xi_{2i}(u)+\xi_{3i}(u)$
reduces to
\begin{eqnarray*}
\xi_i(u)&=&\frac{1}{S(t_1)-S(t_2)}\int_{t_1}^{t_2}\frac{S(s)
M^V_i(ds,u)}{G(s)}\\
&&{}-\frac{1}{(S(t_1)-S(t_2))^2}\int_{t_1}^{t_2}\frac{H_{t_1,t_2}(s,u)M_i(ds)}{G(s)}.
\end{eqnarray*}

Since $\xi_i(u)$ can be written as sums and products of monotone
functions of $u$, therefore, $\{\xi_i(u)\}$ forms a manageable
sequence [Pollard (\citeyear{P1990}), Bilias, Gu and Ying (\citeyear{BGY1997})].  The weak
convergence of $\sqrt{n}(\hat{\mu}_{t_1,t_2}(u)-{\mu}_{t_1,t_2}(u))$
follows from the functional central limit theorem [Pollard (\citeyear{P1990})].
\end{appendix}

\section*{Acknowledgments}
The content of this article is based on the first author's Ph.D.
dissertation conducted at Johns Hopkins University under the
supervision of the second author.  The authors would also like to thank
Professor Norman Breslow for suggestions that improved the presentation
of this paper.

\vspace*{-1,5pt}

\printaddresses

\end{document}